\documentclass[dvips]{article}

\usepackage{amssymb, amsmath}
\usepackage{delarray}
\usepackage{graphicx}
\usepackage{icrc2011}

\title{High-energy spectrum and zenith-angle distribution of atmospheric neutrinos}
\newcommand{\etal}{\MakeLowercase{\textit{et al. }}} % "et al."
\shorttitle{S.Sinegovsky \etal High-energy spectrum of atmospheric neutrinos}

\authors{S. I. Sinegovsky$^{1}$, O. N. Petrova$^{1}$, T. S. Sinegovskaya$^{2}$ }
\afiliations{$^1$Irkutsk State University, Irkutsk, 664003, Russia \\
 $^2$Irkutsk State Railway University, Irkutsk, 664046, Russia}
\email{sinegovsky@api.isu.ru}

\abstract{High-energy neutrinos, arising from decays of mesons produced through the collisions of cosmic ray particles with air nuclei, form the background in the astrophysical neutrino detection problem.  An ambiguity in high-energy behavior of pion and especially kaon production cross sections for nucleon-nucleus collisions
may affect essentially the calculated neutrino flux. We present results of the calculation of the  energy spectrum and zenith-angle distribution of the muon and electron atmospheric neutrinos in the energy range 10 GeV  to 10 PeV.  The calculation was performed  with usage of  known hadronic models (QGSJET-II, SIBYLL 2.1, Kimel \& Mokhov) for two of the primary spectrum parametrizations, by Gaisser \& Honda and by Zatsepin \& Sokolskaya. 
The comparison of zenith angle-averaged muon neutrino spectrum with the measurement data in IceCube experiment make it clear that even at energies above 100 TeV the prompt neutrino contribution is not so apparent because of tangled uncertainties of the strange (kaons) and charm ($D$-mesons) particle production cross sections. An analytic description of calculated neutrino fluxes is presented.
}


\begin{document}
\maketitle

\section{Introduction}

Atmospheric neutrinos (AN) appear in decays of mesons (charged pions, kaons etc.) produced through collisions of  high-energy cosmic rays with air nuclei.  The AN flux in the wide energy range remains the issue of the interest  since the low energy AN flux is a research matter in the neutrino  oscillations studies, and the high energy atmospheric neutrino flux is now appearing as the background noise  for astrophysical neutrino experiments~\cite{frejus,nt200-11,nt200-10,amanda10,icecube11,icecube11_mont,antares11}. 
 
In spite of numerous AN flux calculations are made (for example~\cite{Volk80, Dedenko89, Lipari93,  nss98, FNV2001,  BGLRS04, HKKM04, kss09}, see  also~\cite{Naumov2001, GH} for a review of 1D and 3D calculations of the  AN flux) there are questions concerning to the flux  uncertainties originated from hadronic interaction models as well as from uncertainties in primary cosmic ray spectra and composition in the ``knee'' region.

In this work we present results of the atmospheric neutrino flux calculation in the range $10$--$10^7$ GeV made with use of the hadronic models  QGSJET-II 03~\cite{qgsjet2}, SIBYLL 2.1~\cite{sibyll} as well as the model by Kimel \&  Mokhov (KM)~\cite{KMN} that were tested also in recent atmospheric muon flux calculations~\cite{kss08, sks10}. We compute here the differential energy spectrum of the conventional neutrinos averaged over zenith angles to compare with the data of the Frejus~\cite{frejus},  AMANDA-II~\cite{amanda10} and IceCube~\cite{icecube11} experiments.

\section{Calculations vs. the experiment \label{sec:results}} 

The calculation is performed on the basis of the  method~\cite{NS} of solution of the hadronic cascade equations in the atmosphere,  which takes into account non-scaling behavior of inclusive particle production cross-sections, the rise of total inelastic hadron-nuclei cross-sections, and the non-power law primary spectrum (see also~\cite{kss09, kss08,  sks10}). Along with major sources of the muon neutrinos, $\pi_{\mu2}$ and  $K_{\mu2}$ decays, we consider three-particle semileptonic decays, $K^{\pm}_{\mu3}$, $K^{0}_{\mu3}$,  the contribution originated from decay chains   $K\rightarrow\pi\rightarrow\nu_\mu$ ($K^0_S\rightarrow \pi^+\pi^-$, $K^\pm \rightarrow \pi^\pm \pi ^0$), as well as small fraction from the muon decays. % $\mu_{e3}$ .
One can neglect the 3D effects in calculations of the atmospheric muon neutrino flux  
near vertical at energies $E \gtrsim 1$ GeV and  at $E \gtrsim 5$ GeV in case of directions close to horizontal (see~\cite{BGLRS04,HKKM04}).
%================================
\begin{figure}[!t]  %\vskip 0.7 cm           %
  \centering
\includegraphics[width=75 mm]{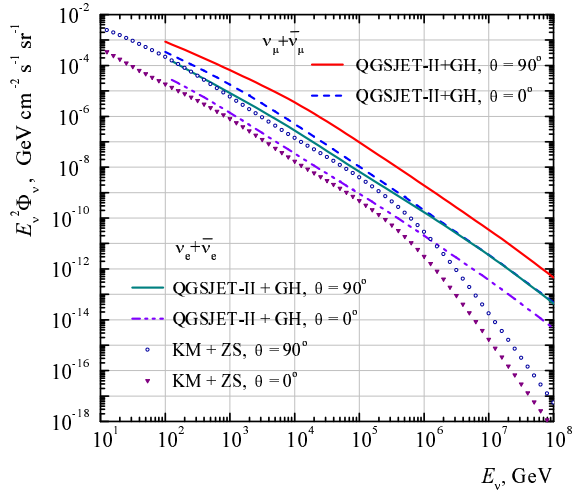} % {nuenumu_sp} 
 \caption{Spectra of the conventional muon and electron neutrinos calculated for vertical and horizontal directions.} 
  \label{nu_numu} 
 \end{figure}
% ================================
%==================================
\begin{figure}[!b]
  \centering \vspace{0.20 cm} % \vskip -1 cm
\includegraphics[ width=70 mm]{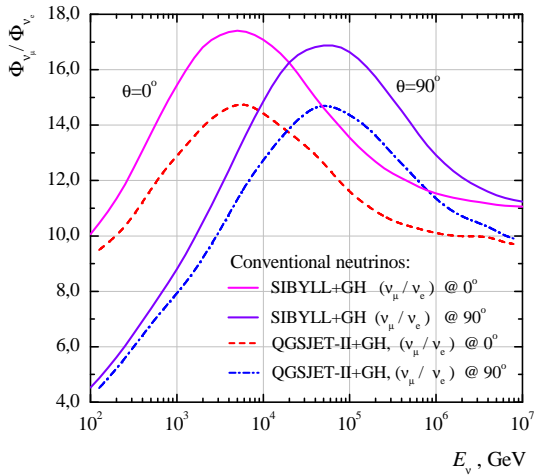} 
 \caption{Conventional neutrino flavor ratios  calculated with SIBYLL and QGSJET-II hadronic models for the GH primary spectrum.} 
\label{numu-nue_rat} %\vskip -0.2 cm
 \end{figure} 
%===================================  
%
 As the primary cosmic ray spectra and composition in wide energy range we use the model recently proposed  by Zatsepin \& Sokolskaya (ZS)~\cite{ZS3C}, which fits well the  ATIC-2 experiment data~\cite{atic2} and supposedly to be valid up to $100$ PeV. The ZS proton spectrum at $E\gtrsim 10^6$ GeV is compatible with KASCADE data~\cite{KASCADE05} as well the helium  one within the range of the KASCADE spectrum obtained with the usage of QGSJET 01 and SIBYLL models. Alternatively in the energy  range $10-10^6$ GeV we use the parameterization by Gaisser, Honda, Lipari and Stanev (GH)~\cite{GH}, the version with the high fit to the helium data. Note this version is consistent with the data of the  KASCADE experiment at $E_0>10^6$ GeV that was obtained (through the EAS simulations) with the SIBYLL 2.1.
%  ================================
\begin{figure}[!ht]            %
  \centering
\includegraphics[width=70 mm]{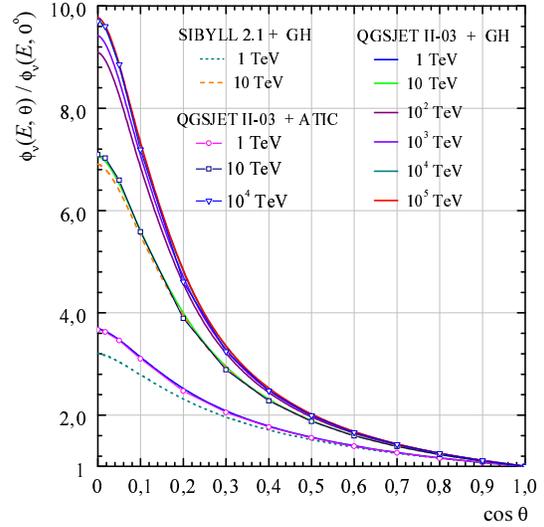} %  2.in
 \caption{Zenith-angle enhancement of the ($\nu_\mu + \bar{\nu}_\mu$) flux.} 
  \label{angle_distrib} %\vskip -0.5 cm
 \end{figure}
%  ================================ 
\begin{figure}[!t]
 \centering %\vskip -1.5 cm
\includegraphics[width=80 mm]{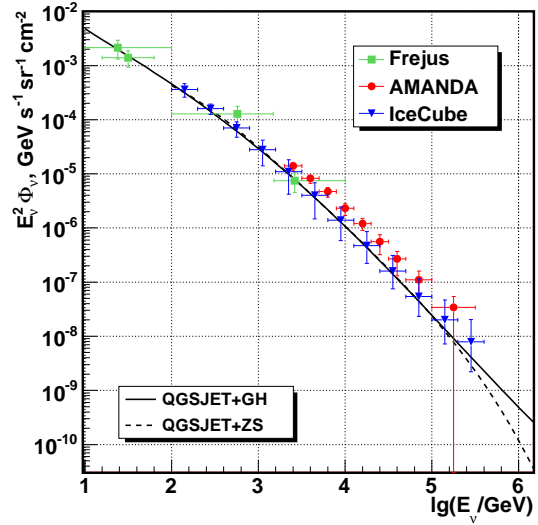} %QGS2+GH+ZS %\vskip -1.5 cm
\caption{Conventional ($\nu_\mu+\bar\nu_\mu$) spectrum averaged over zenith angles. Curves: the calculation with usage of  QGSJET-II. Symbols: data of experiments, Frejus~\cite{frejus}, AMANDA-II~\cite{amanda10}, IceCube~\cite{icecube11}.}
\label{qgs_2pcr}  
\end{figure}
%====================================
%====================================
\begin{figure}[!t]     
\centering
\includegraphics[width=80 mm]{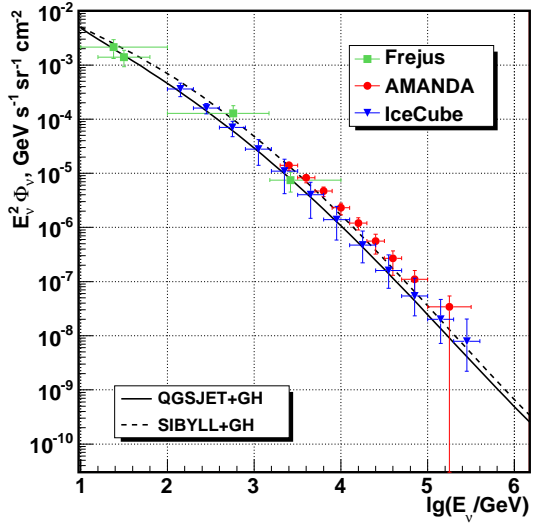}%2mod+GH_10 %{SIBYLL+GH+ZS} 
\caption{Comparison of the ($\nu_\mu+\bar\nu_\mu$) spectrum calculations with QGSJET-II and  SIBYLL 2.1}
\label{2mod_gh-10GeV}
%\label{sib_2pcr}  
\end{figure}
%====================================

 Figures~\ref{nu_numu}, \ref{numu-nue_rat} display the scale of difference between  the conventional ($\nu_\mu+\bar\nu_\mu$) spectra and  ($\nu_e+\bar\nu_e$) one, calculated with usage of QGSJET{\text -}II, SIBYLL and KM model for GH and ZS primary spectra. The difference of neutrino flux predictions related to choice of hadronic models is clearly apparent. %primary cosmic ray spectra, GH and ZS,

Zenith-angle distributions of the conventional neutrinos, $\phi_{\nu_{\mu}}(E,\theta)/\phi_{\nu_{\mu}}(E,0^\circ)$, for the energy range $1$-$10^5$ TeV  are shown in Fig.~\ref{angle_distrib}. Calculations are made with  QGSJET{\text -}II and SiBYLL 2.1 models both for GH  and ZS (ATIC-2) primary spectra and composition. 
As was expected, a shape of the angle distribution visibly depends on the neutrino energy (at  $E < 100$ TeV) especially for directions close to horizontal. The effect of the hadronic models (as well as of the primary spectrum) on the angle distribution is rather weak, while the spectra differences amount to $80\%$~\cite{kss09}.

The calculation of the conventional ($\nu_\mu+\bar\nu_\mu$)  flux averaged over zenith angles
as compared with Frejus~\cite{frejus} (squares), AMANDA-II~\cite{amanda10} (circles), and IceCube~\cite{icecube11} (triangles) measurement data is shown in Figs.~\ref{qgs_2pcr}, \ref{2mod_gh-10GeV}.  Figure~\ref{qgs_2pcr}  displays  the conventional ($\nu_\mu+\bar\nu_\mu$)  spectrum (averaged over zenith angles in the range $0^\circ-84^\circ$)  calculated with usage of QGSJET{\text -}II model  for GH primary spectra and composition (solid line) as well as for ZS one (dashed). 
 The difference in neutrino flux predictions resulted from the primary cosmic ray spectra becomes apparent 
 at high neutrino energies: the flux obtained with QGSJET{\text -}II for  GH spectrum at $600$ TeV is  nearly twice as large as that for ZS spectrum. At 1 PeV  this discrepancy increases to the factor about five. 
Comparison of QGSJET{\text -}II and SIBYLL  presented in Fig.~\ref{2mod_gh-10GeV} shows that the former seems more preferable to desribe the IceCube measurements at the  energies below $40$ TeV (conventional neutrinos).
   
The usage of QGSJET-II and SIBYLL models leads to apparent difference in the neutrino flux, as well as in the case of SIBYLL as compared to KM  (unlike the muon flux, where SIBYLL and KM lead to very similar results ~\cite{kss08}). On the contrary, the QGSJET-II neutrino flux is very close to the KM one: up to $100$ TeV the difference does not exceed $5\%$ for the GH spectrum and $10\%$  for the  ZS one at $\theta=0^\circ$. 
Note that the muon flux discrepancy in QGSJET-II and KM predictions is about $30\%$ at vertical~\cite{kss08}. 
%=================================
 \begin{figure}[!t]
  \centering %\vspace{0.20 cm} % \vskip -1 cm
\includegraphics[width=80 mm]{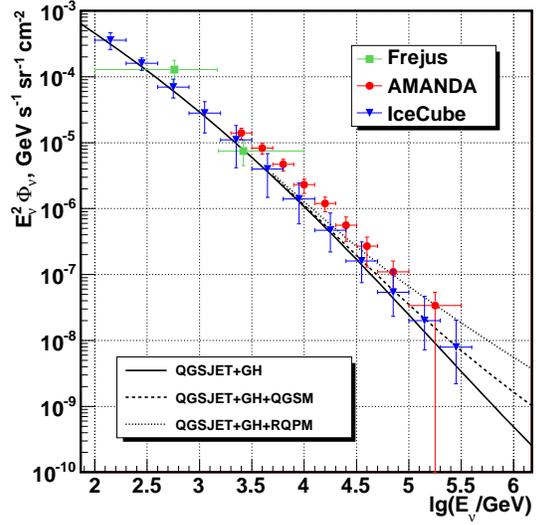} % QGSJET+GH10+pr %{fig04}   {QGSJET+GH+pr} - from 100 GeV
 \caption{Flux of the conventional and prompt muon neutrinos (case of GH spectrum).} 
\label{qgs_gh_pr} %\vskip -0.2 cm
 \end{figure} 
%=================================== 
%=================================
 \begin{figure}[!th]
  \centering %\vspace{0.20 cm} % \vskip -1 cm
\includegraphics[width=80 mm]{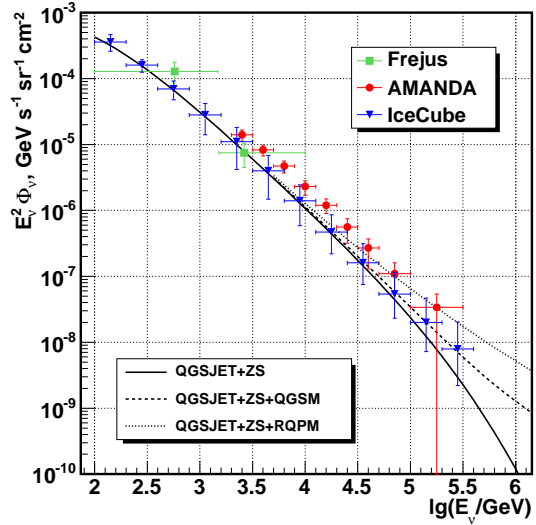}%{QGSJET+ZS+pr%} {fig04}   {QGSJET+ATIC_num_aver}
 \caption{Flux of the conventional and prompt muon neutrinos (case of QGSJET-II +  ZS spectrum).} 
\label{qgs_zs_pr} %\vskip -0.2 cm
 \end{figure} 
  
 Figure~\ref{qgs_gh_pr}  shows the sum of the conventional flux (calculated for GH spectrum with usage of   QGSJET-II) and  prompt muon neutrino flux  predictions~\cite{bnsz89} (see also~\cite{nss98, Naumov2001, prd98}) due to  nonperturbative  models, the recombination quark-parton model (RQPM, dotted line) and  the quark-gluon string model (QGSM, dashed line).   The case of ZS spectrum one can see in Fig.~\ref{qgs_zs_pr}.  
 The prompt neutrino fluxes were obtained~\cite{bnsz89}  with NSU primary spectrum~\cite{NSU}, therefore they can serve here as upper limits for the prompt neutrino flux due to RQPM  and  QGSM.  
%Notice  
 It worth noting that  evaluation of the prompt neutrino flux obtained with the dipole model~\cite{DM}  is close to the QGSM prediction~\cite{bnsz89} above  $1$ PeV. 
The prompt neutrino flux due to QGSM in the energy range $5$ TeV $\leq E_{\nu} \leq 5\cdot 10^3$ TeV was approximated by the expression 
\begin{equation}\label{pms}
\Phi_{\nu}^{\rm pr}(E_{\nu})
 = A(E_1/E_\nu)^{3.01}[1+(E_1/E_\nu)^{2.01}]^{-0.165}, 
\end{equation}
 where $A=1.19\cdot 10^{-18}\,\mathrm{(GeV\,cm^{2}\,s\,sr)^{-1}}$, $E_1=100$ TeV.  In this range we neglect the  weak angle dependence of the prompt neutrino flux.  
 
\section{Approximation formula}

Numerical results of the conventional muon neutrino spectra in the energy range $10^2-10^6$ GeV for different zenith angles can be approximated with accuracy ($3\div 8$)$\%$ by the formula:
\begin{equation}
\log_{10}[E_{\nu}^{2}\Phi_{\nu}(E_{\nu},\theta)]=\sum_{k=0}^{4}\sum_{n=0}^3 a_{kn}x^ny^k.	
\end{equation}
Here $\Phi_{\nu}(E_{\nu},\theta)$ is the flux with units of GeV$^{-1}$ s$^{-1}$ sr$^{-1}$ cm$^{-2}$, $x=\cos\theta$, 
 $y=\log_{10}(E_{\nu}/\rm{GeV})$. Coefficients $a_{kn}$ are given in Tables~\ref{tab_qgs+gh}, \ref{tab_qgs+zs}. 

\begin{table}[!ht]
\begin{center} 
\caption{Coefficients $a_{kn}$ for the QGSJET+GH choice} \vskip 0.20 cm%\vskip -1.0 cm 
\begin{tabular}{|c|c|c|c|c|}
\hline
  & \multicolumn{4}{|c|}{$n$} \\ \cline{2-5}
$k$	&$0$	&$1$	&$2$	&$3$ \\
\hline
 & \multicolumn{4}{|c|}{$a_{kn}$} \\ %\hline
%\cline{1-1}
$0$	& 0.72550	&-6.45625	& 4.10284 	&-0.87026 \\
$1$	&-3.21166	& 6.38522	&-3.31293 	& 0.38300 \\
$2$	& 1.00337	&-2.46611	& 0.99745   & 0.01675 \\
$3$	&-0.19397	& 0.35758	&-0.07515 	&-0.04540 \\
$4$	& 0.01211 &-0.01753	&-0.00135	  &0.00537 \\
\hline\end{tabular}
\label{tab_qgs+gh}
\end{center} %\vskip -0.7 cm
\end{table}
%-------------------------------
%==================================
\begin{table}[!ht]
\begin{center}
\caption{Coefficients $a_{kn}$ for the  QGSJET+ZS choice} \vskip 0.20 cm
\begin{tabular}{|c|c|c|c|c|}
\hline
 & \multicolumn{4}{|c|}{$n$} \\ \cline{2-5}
 $k$	&$0$	&$1$	&$2$	&$3$ \\
\hline
 & \multicolumn{4}{|c|}{$a_{kn}$} \\ %\hline
$0$	&-3.21881	&-7.00088	& 4.64475	&-1.07882 \\
$1$	& 1.60632	& 6.92858	&-3.87209	& 0.60250 \\
$2$	&-1.11835 &-2.65863	& 1.20056	&-0.06412 \\
$3$	& 0.20848 & 0.38659	&-0.10641	&-0.03279 \\
$4$	&-0.01577	&-0.01909	& 0.00037	& 0.00467 \\
\hline
\end{tabular}
\label{tab_qgs+zs}
\end{center} %\vskip -0.7 cm
\end{table}
%==========================================

\section{Summary} 

 The calculations of the high-energy atmospheric muon neutrino flux demonstrate rather weak dependence on the primary specrtum models in the energy range $10-10^5$ GeV.  However the picture appears less steady because of sizable flux differences originated from  the models of high-energy hadronic interactions. As it can be seen by the example of the models QGSJET-II and SIBYLL 2.1, the major factor of the discrepancy in the conventional neutrino flux is the kaon production in nucleon-nucleus collisions.  
 
Comparison of calculated muon neutrino flux  with the spectrum measured by IceCube shows that QGSJET-II is preferable
model irrespective of the primary spectrum choice. The prompt neutrino contribution due to quark qluon string model (QGSM) added to the conventional one lead to better agreement with the IceCube measurement above 100 TeV.
 
%\begin{acknowledgments}
The work was supported by Russian Federation Ministry of Education and Science within the Federal  Programs "Scientific and educational specialists for innovative Russia" under Contracts 14.740.11.0890, P681, P1242, and  "Development of scientific potential in Higher Schools" under grant 2.2.1.1/12360. 
%\end{acknowledgments}

%\bigskip

\end{document}